\documentclass[twocolumn,showpacs,preprintnumbers,amsmath,amssymb]{revtex4}
\usepackage{graphicx}
\usepackage{amsmath}
\usepackage{bm}% bold math

 \newcommand{\iu}{\textrm{i}}

\begin{document}

\title{Stable multiple-charged localized optical vortices in
cubic-quintic nonlinear media}

\author{T.A. Davydova}
% \email{tdavyd@kinr.kiev.ua}
\author{A.I. Yakimenko}%
\email{ayakim@kinr.kiev.ua}

%A.I. Yakimenko\footnote[3]{To whom correspondence should be
%addressed}}% (ayakim@kinr.kiev.ua)} }

\address{Institute for Nuclear Research, Prospect Nauki 47,
Kiev 03680, Ukraine}
%%% ----------------------------------------------------------------------
\date{\today}
%-----------------------
\begin{abstract}
The stability of two-dimensional bright vortex solitons in a media
with focusing cubic and defocusing quintic nonlinearities is
investigated analytically and numerically. It is proved that
above some critical beam powers not only one- and two-charged but
also multiple-charged stable vortex solitons do exist. A vortex
soliton occurs robust with respect to symmetry-breaking
modulational instability in the self-defocusing regime provided that
its radial profile becomes flattened, so that a self-trapped wave
beam gets a pronounced surface. It is demonstrated that the
dynamics of a slightly perturbed stable vortex soliton resembles
an oscillation of a liquid stream having a surface tension. Using
the idea of sustaining effective surface tension
for spatial vortex soliton in a media with competing nonlinearities the
explanation of a suppression of the modulational instability is proposed.
\end{abstract}
\maketitle
%%% ----------------------------------------------------------------------
\section{Introduction}
%%% ----------------------------------------------------------------------
Spatial optical solitons and vortex solitons are self-trapped
light beams of finite cross section that are supported by a
balance between linear diffraction (dispersion) and nonlinear
self-focusing of the intense wave. These structures have been
predicted and experimentally demonstrated in various nonlinear
optical media (see \cite{KivsharAgrawal} and references therein).
An optical vortex soliton has embedded phase singularity and
carries intrinsic angular momentum. The phase circulation around
the axis of propagation is equal to $2\pi m$, where integer
number $m$ is the topological charge of the vortex. We
investigate the possibility of formation and stability of
two-dimensional multiple-charged (having topological charge $m$
to be equal up to five) localized envelope vortices on the basis
of the Nonlinear Schr\"{o}dinger Equation (NSE) with competing
cubic-quintic (CQ) nonlinearities.

It is well known, that ordinary nonspining bright solitons are
stable in CQ nonlinear media. Though all localized optical vortex
solitons (LOVS) were believed  \cite{Skryabin98} to possess a
strong azimuthal modulation instability. As a result, an unstable
vortex decays into several spatial solitons with zero topological
charges, which fly off tangentially to the initial vortex ring
conserving the total angular momentum. Recent investigations
\cite{Berezhiani:01,TowersPLA02,Malomed02,OurPRE03} have shown,
however, that in CQ nonlinear media one- and two-charged LOVS
become stable if their numbers of quanta (beam power) exceed the
critical values $N_{\textrm{cr}}$. Nevertheless, vortices with
$m>2$ were still regarded to be unstable because of the modulation
instability \cite{TowersPLA02}. We show here that, in contrast
with what was previously believed, even higher-order (with $m>2$)
vortices become stable above certain threshold value
$N_{\textrm{cr}}$, where $N_{\textrm{cr}}$ increases with
topological charge.

The important, but still open question is: what is the physical
essence of a vortex ($m\ne 0$) symmetry-breaking instability
suppression? It is remarkable, that in self-defocusing regime
(when an increase of the input power leads to broadening of the
light beam) vortex as well as soliton shape changes abruptly
above certain critical power: the radial intensity distributions
becomes practically uniform with pronounced surface. As was
demonstrated previously \cite{OurPRE03}, this modification of the
soliton profile corresponds to some bifurcation. It is
interesting that here the light beam gets some common features
with a liquid stream. Actually, the sharpness of the soliton
shape modification corroborates the conception \cite{Teixeiro:02}
of "phase gas-liquid transition". The authors of Ref.
\cite{Teixeiro:02} demonstrated that a laser light gets some kind
of surface tension above certain threshold beam power (in a
``liquid" state). By numerical simulations of soliton collisions
against planar boundaries and localized inhomogeneities they
proved that two-dimensional ``liquid solitons" behave like liquid
streams having a surface tension.

It is reasonable to expect that such ``gas-liquid transitions" do
happen with vortex solitons too.  In this paper we propose and
validate the explanation of a vortex soliton stabilization on the
basis of the sustaining surface tension conception. We show here
that the dynamics of slightly perturbed stable vortex resembles
the oscillation of a liquid stream having a surface tension. The
remarkable analogy of light ``condensation" with an effect of
condensate droplet formation \cite{KovalevKosevich76} in the
$N$-particle boson system at the large $N$ limit is revealed.
%%% ----------------------------------------------------------------------
\section{Vortex solitons}\label{sec:vortexGENERAL}
%%% ----------------------------------------------------------------------
We consider here the intense laser light beam propagating in CQ
media with focusing cubic and defocusing quintic nonlinearities.
Refractive index of CQ materials can be approximated as follows:
$n=n_0+n_2I-n_4I^2$, here $I$ is the intensity of light beam,
coefficients $n_2$, $n_4>0$ determine nonlinear response of the
media. It is important to note, that for some nonlinear materials
\cite{Teixeiro:02} this fitting formula works even in strongly
nonlinear self-defocusing regime, when $I>n_2/2n_4$, so that
$\partial n/\partial I<0$.

The electromagnetic field envelope $\Psi(x,y,z)$ of linearly
polarized laser beam, which propagates along
 $z$ through a CQ nonlinear optical material, is described
  in the paraxial approximation by
generalized NSE (GNSE):
\begin{equation}
\label{eq:GNSE} \iu\partial_z\Psi + D\Delta_{\perp} \Psi +B\left|
\Psi \right|^2\Psi -K\left| \Psi \right|^4\Psi= 0,
\end{equation}
where $\Delta_{\perp}=\partial^2/\partial x^2+\partial^2/\partial
y^2$ is the Laplacian operator, $D=1/2\kappa n_0,$ $B=\kappa
n_2,$ $K=\kappa n_4,$ $\kappa$ is the wave number of the light
beam.

\begin{figure}[e]
\begin{centering}
\includegraphics[width=8.5cm]{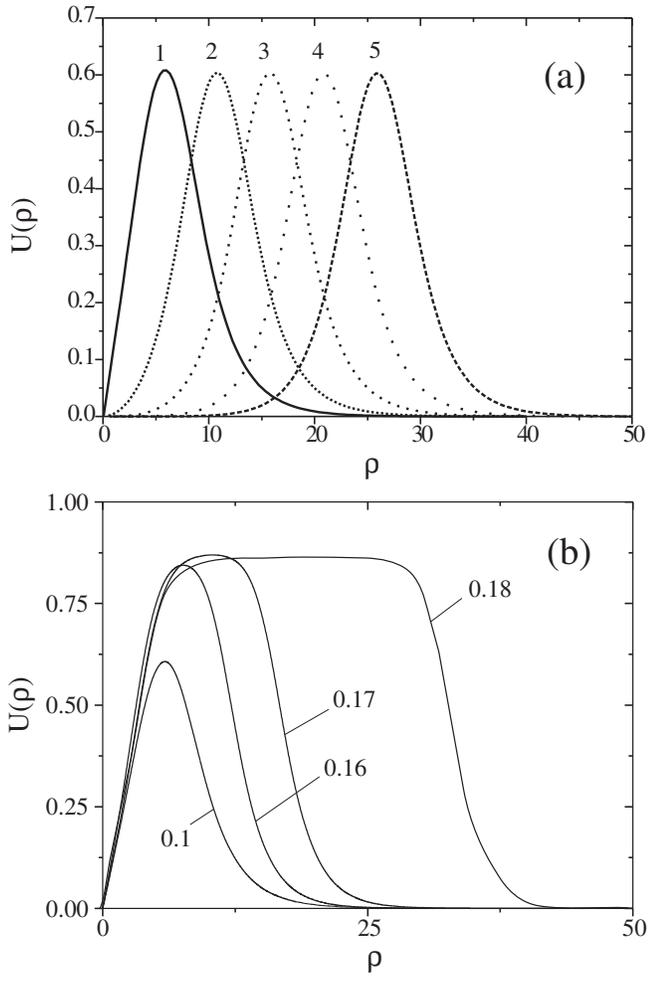}
 \caption{Radial profiles of vortex solitons: (a) $m=1,..,5$ ($\lambda=0.1$),
 (b) the one-charged vortices with different $\lambda$.}
  \label{Cuts}
\end{centering}
\end{figure}

For localized solution of the GNSE (\ref{eq:GNSE}) the following
integrals of motion are assumed to be finite: (i) beam power (or
number of quanta): $N = \int\left| \Psi \right|^2 d^2 \textbf{r}$,
%%%%%
(ii) momentum: $\vec{P} = \int \vec p \,d^2 \textbf{r},$ where $
\vec p=-\frac{\textrm{i}}{2}\left( \Psi^* \nabla \Psi - \Psi
\nabla \Psi^* \right),$
%%%%%
(iii) angular momentum: $\vec{ M }= \int \left[\vec r \times \vec
p\right] d^2 \textbf{r},$ (iv) Hamiltonian: $H =\int h\, d^2
\textbf{r},$ where $h=D|\nabla \Psi|^2-\frac{1}{2}B|\Psi|^4+
  \frac{1}{3}K|\Psi|^6.$
%%% ----------------------------------------------------------------------
Vortex solitons are stationary solutions of the GNSE
(\ref{eq:GNSE}) of the form
\begin{equation}\label{eq:vars_separ}
\Psi(\textbf{r}, z)=\psi(r)e^{\iu m\varphi+\iu\Lambda z},
\end{equation}
where $m$ is the topological charge and $\Lambda$ is the
propagation constant. The radial profile $\psi(r)$ of the
localized vortex soliton can be found by means of numerical
solution of the ordinary differential equation:
\begin{equation}
 \label{eq:GNSE2D_unitless}
-\lambda U + \Delta_\rho^{(m)}U +U^3 -U^5 = 0,
\end{equation}
where $$\Delta_\rho^{(m)}=\frac{d^2 }{d \rho^2}
+\frac{1}{\rho}\frac{d }{d \rho}-\frac{m^2}{\rho^2},$$ and we have
introduced the following rescaling transformation:
$\rho=r\sqrt{B^2/DK},$ $\lambda=\Lambda K/B^2,$
$U(\rho)=\psi(r)\sqrt{K/B}$. The typical profiles of vortices with
different topological charges are presented in the Fig.
\ref{Cuts} (a). Each value of the propagation constant $\lambda$
corresponds to vortex soliton solution having definite beam power
$N(\lambda).$

%-----------------------------------------------
The main properties of the stationary solutions can be
investigated by approximate variational method. The variational
analysis for ordinary nonspining solitons, which takes into
account soliton shape modification has been performed in Ref.
\cite{OurPRE03,SuperGauss98}. Here we restrict variational
analysis of the LOVS using the trial function with unchanged
radial profile:
$$\Psi(r, z)=
A(z)r^{|m|}\exp\left\{-\frac12\left[r/a(z)\right]^2+\iu
\gamma(z)r^2+\iu m\varphi\right\},$$ where $\gamma(z)$ is the
phase curvature, parameters $a(z)$ and $A(z)$ characterize
beam width and amplitude respectively. The parameter $A(z)$ can be
expressed in terms of $N$ and $a(z)$ using definition of the
number of quanta $N$.
 After the procedure of Ritz
optimization one can obtain for the variables $a(z)$ and
$\beta(z)=\gamma(z) a(z)$ the set of equations in canonical form,
which describes the evolution (in $z$-direction) of the vortex
parameters:
\begin{eqnarray}\label{eq:HamiltDynamSet1}
\frac{N(m+1)}{2}\frac{d a}{d z}=\frac{\partial
H}{\partial\beta},\\
\label{eq:HamiltDynamSet2} \frac{N(m+1)}{2}\frac{d \beta}{d
z}=-\frac{\partial H}{\partial a}.
\end{eqnarray}The Hamiltonian in the variational approximation is
given as follows:
\begin{equation}\label{eq:HamiltVariat}
H=N\left\{D(m+1)(1/a^2+\beta^2)-\frac{1}{2}\frac{bN}{a^2}
+\frac{1}{3}\frac{kN^2}{a^4}\right\},
\end{equation}
where
$$b=\frac{B(2m)!}{\pi(m!)^22^{2m+1}}, k=\frac{K(3m)!}{\pi^2(m!)^33^{3m+1}}.$$
The vortex soliton corresponds to the stationary point of the set
(\ref{eq:HamiltDynamSet1}), (\ref{eq:HamiltDynamSet2}):
\begin{equation}\label{eq:stationary_condit}
 \frac{\partial H}{\partial a }=0,
\, \frac{\partial H}{\partial\beta }=0,
\end{equation}
which determine parameters $a$ and $\beta$ of a vortex:
\begin{equation}\label{eq:a_0beta_0}
a_0^2=\frac{4k}{3b}\frac{N^2}{N-N_m}, \, \, \beta_0=0,
\end{equation}
 where $$N_m=2D(m+1)/b.$$
From Eq. (\ref{eq:a_0beta_0}) it follows that (i) $m$-charged
vortex soliton exists only above the threshold beam power:
$N>N_m$, (ii) if beam power exceeds the doubled threshold value:
$N>2N_m$, the self-focusing regime turns into the self-defocusing
one, (iii) for $N\gg N_m$ the effective radius $a_0$ of the
two-dimensional structure has the asymptotic behaviour of the
form: $a_0\sim N^{1/2}$. As was found in Ref. \cite{OurPRE03}, the
radial profile of an ordinary soliton in self-defocusing regime
changes abruptly, if beam power exceeds quadruple threshold power
for soliton existence: $N>4N_0$. It is remarkable, that the point
of ``gas-liquid" condensation of nonspining light beam, which has
been revealed in Ref. \cite{Teixeiro:02}, also corresponds to
power $N=4N_0$: laser light with $N>4N_0$ attains some kind of
effective surface tension. Above certain value of the beam power
a vortex soliton gets a sharp boundary too [see Fig. \ref{Cuts}
(b)], likewise the gas cloud, which condenses into the liquid
droplet. This state with nearly uniform density of quanta $\sim
|\Psi|^2$ can be considered as a condensate stream flowing along
the axis $z$.

A deeper insight into physical essence of this phenomena can be
obtained using the analogy between the models, describing the
photon gas (\ref{eq:GNSE}) and bosons interacting via
$\delta$-like potential. In Ref. \cite{KovalevKosevich76} the
one-dimensional system of a finite number of bosons with the
pairwise attraction and three-body repulsion has been considered
in the self-consistent field approximation on the basis of the
one-dimensional NSE with CQ nonlinearity. As was demonstrated in
Ref. \cite{KovalevKosevich76}, an addition of particles to such
boson system causes the increase of the density of the particles
until the three-body repulsion becomes significant. The further
increase of $N$ leads to appearance of the state having nearly
homogeneous distribution of the particles and constant binding
energy. This $N$-particle bound state in the large $N$ limit has
been considered as condensate droplet in the coordinate space
\cite{KovalevKosevich76}. The ``trapping energy" per one particle
(the Hamiltonian per one quantum) tends to the finite negative
value, for two-dimensional LOVS under consideration, as it had
been found for self-trapped bosons \cite{KovalevKosevich76}:
$$\varepsilon=\lim_{N\to\infty}H/N=-\frac{3b^2}{16|k|}.$$
It is clear that binding energy per unit length along $z$ (the
Hamiltonian $H$) for the finite size system is less than it would
be for the infinite one, since not all ``binding interreactions"
act on ``surface particles". This leads to appearance of a surface
tension and negative surface energy. As was shown in
\cite{KovalevKosevich76}, the value
$$H_s=\lim_{N\to\infty}(H-N\varepsilon)=-2N_m\varepsilon,$$
can be treated as an analog of the surface power. In the next
section we investigate how the surface power, which appears as an
effective surface tension, influences on the vortex structure
evolution in $z$-direction.
%-------------------------------------------------
\section{ Dynamics of perturbed vortex soliton}
\begin{figure}[e]
\begin{centering}
\includegraphics[width=8.5cm]{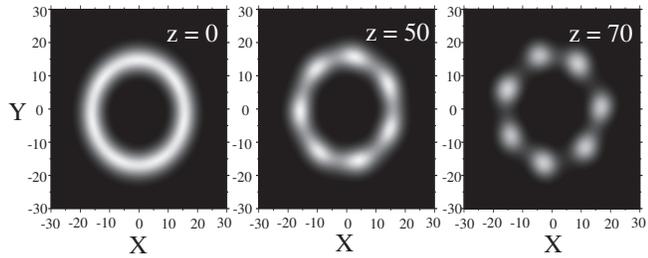}
 \caption{Gray-scale intensity distribution
 of vortex wave beam propagating in $z$-direction.
 The unstable vortex soliton with
 $m=3$, $\lambda=0.1$ decays into ordinary solitons.}
  \label{Decays}
\end{centering}
\end{figure}

%-------------------------------------------------
As known, in contrast to nonspining solitons, which are stable
with respect to rather large symmetric and asymmetric
perturbations, LOVS are stable for small radially-symmetric
perturbation, but unstable with respect to asymmetric ones. A
small azimuthal perturbation may grow forming several filaments,
and, as the result, a LOVS decays into a few ordinary solitons,
which fly off in such a way, that the total angular momentum of
the system conserves. A typical example of the dynamics (in
$z$-direction) of unstable vortex is drawn in Fig. \ref{Decays}.

\begin{figure}[e]
\begin{centering}
\includegraphics[width=8.5cm]{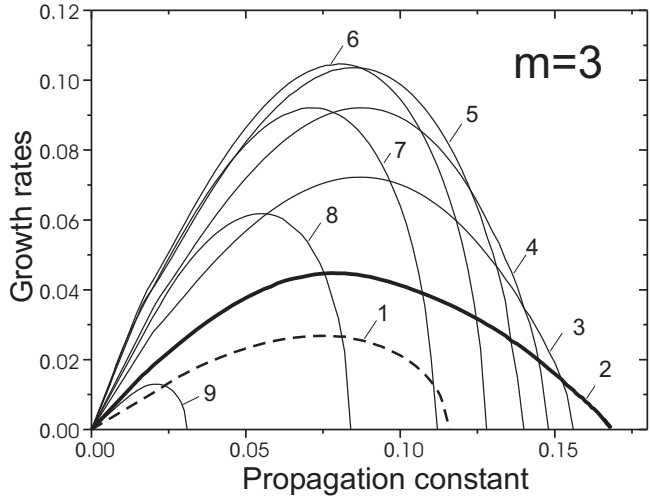}
 \caption{Maximum growth rates of all unstable azimuthal
eigenmodes vs propagation constant $\lambda$ for vortex solitons
with $m=3$. Integers near the curves indicate  azimuthal numbers
$L$.}
  \label{fig:GRrate_BDLE0}
\end{centering}
\end{figure}

To consider an initial stage of the instability,
we have performed the linear stability analysis of the LOVS
with respect to small perturbations:
$$
\Psi(\vec{r}, z)=\left\{\psi(r)+\varepsilon (\vec r,z)\right\}
\exp\left\{\iu m\varphi+\iu\Lambda z\right\}
$$
with different azimuthal periods:
$$\varepsilon (\vec r,z)=a^+(r,z)e^{\iu L\varphi}
+a^-(r,z)e^{-\iu L\varphi},$$ where $|\varepsilon (\vec r,z)|\ll
|\psi(r)|$. The  azimuthal number of the perturbation $L=0, 1,
2,... $ determines the number of humps on the envelope surface.
Since we investigate the linear stage of instability, all unstable
modes may be considered independently and treated as
exponentially growing: $\varepsilon (\vec r,z)\sim e^{\Gamma_L
z}.$ Therefore, the linear stability analysis after the
linearization is reduced to solving the eigenvalue problem for
growth rates $\Gamma_L$. If $\Gamma_L>0$, the perturbation grows
up, so that LOVS would decay into $L$ filaments. We have solved
this eigenvalue problem numerically. The growth rates of the all
unstable eigenmodes for LOVS with $m=3$ are presented in the Fig.
\ref{fig:GRrate_BDLE0}. It is seen that the widest instability
region corresponds to $L=2$ and the higher-order perturbation are
suppressed even for small $\Lambda$ in the vicinity of the
threshold $N_m$. As seen from Fig. \ref{fig:GRrate_BDLE0},  above
some critical beam power $N_{\textrm{cr}}$ the modulation
instability of LOVS with $m=3$
 is completely
suppressed -- all growth rates are equal to zero. The same
analysis has been performed for higher-order LOVS (up to $m=5$),
and it has been found that above some critical beam powers
$N>N_{\textrm{cr}}$ multiple-charged LOVS are stable.

The LOVS stability has been confirmed by means of direct numerical
simulation of the nonstationary GNSE (\ref{eq:GNSE}) with the
perturbed LOVS as initial condition. The dynamics of stable LOVS
has been found to be quasiperiodical -- effective width and
amplitude of perturbed vortex oscillate with $z$.

Let us consider the dynamics of small oscillations of a LOVS
surface in more details. In the framework of variational analysis
the parameters of the stationary LOVS are determined by
(\ref{eq:a_0beta_0}). Expanding the Hamiltonian around the
stationary point one straightforwardly obtains that dynamics of
effective width $a(z)$ of a vortex wave beam is described by the
Newton-like motion equation:
$$\mu \frac{d^2 \xi}{d z^2}=-\xi \left(\frac{\partial^2 H}{\partial a^2}
\right)_{\beta=0,\, a=a_0},$$ where $\xi(z)=a(z)-a_0$,
$\mu(N)=N(m+1)/8D.$ Thus, the frequency of small oscillations of
slightly perturbed vortex soliton is:
\begin{equation}\label{eq:frequency}
\omega^2=\frac{4}{a_0^2}\frac{H_s}{\mu_0}\left(1-\frac{N_m}{N}\right)^2,
\end{equation}
where $\mu_0=\mu(N_m)$, $H_s$ is the surface power. If $N\gg N_m$,
the frequency tends to $\omega\to\sqrt{4H_s/a_0^2\mu_0}$. Hence,
the dynamics of perturbed vortex soliton at large $N$ limit looks
like oscillations of liquid stream having the effective surface
tension
 $\sigma\sim H_s/2\pi a_0$ and the effective density
$\rho_0\sim\mu_0/\pi a_0^2$.

In conclusion, we have found the conditions for  the existence of
stable multiple-charged localized optical vortex solitons in a CQ
nonlinear media. They occur stable above some critical beam powers
$N_{\textrm{cr}}$, in the self-defocusing regime provided that
their profiles become flat-topped. The increase of LOVS beam power
above $N_{\textrm{cr}}$ leads to formation of the vortex
structure having the pronounced plateau on the radial intensity
distribution and the sharp boundary. We have found out that the
dynamics of the slightly perturbed stable vortex is similar to
oscillations of liquid stream having a surface tension. We have
drawn a parallel between the known phenomena of condensation in
the boson system and the laser light beam ``condensation". We
have estimated the surface beam power and corroborated that LOVS
has an effective surface tension, which causes a vortex
stabilization. Actually, the sustaining surface tension
suppresses small modulations of the vortex surface. It impedes the
formation of the filaments, therefore LOVS becomes stable in a
``liquid" state.

We believe that described mechanism can be used to explain LOVS
stabilization in other materials with competing
focusing-defocusing nonlinearities, e.g. in the media with
defocusing cubic and focusing quadratic ones
\cite{MihalachePRE02}. Moreover, the insight into the physical
reasons of LOVS stabilization would allow one to choose more
suitable nonlinear material for experimental realization of the
stable vortex solitons.
%---------------------------------------------
\bibliography{RefsJOptA}
%---------------------------------------------
%--------------------
%\begin{figure}[htb]
%\begin{centering}
%\includegraphics[width=8.5cm]{CutsM1.eps}
% \caption{Radial profiles of the one-charged vortex with different $\lambda$.}
%  \label{CutsM1}
%\end{centering}
%\end{figure}
%--------------------
%--------------------
%=========================================
\end{document}